\documentclass[12pt,a4paper,twocolumn]{scrartcl}

\usepackage[utf8]{inputenc}

\usepackage[english]{babel}

\usepackage{amsmath}

\usepackage{microtype}

\usepackage{csquotes}

\usepackage[top=2.5cm, bottom=2.5cm, left=1.8cm, right=1.8cm]{geometry}
\usepackage[font=footnotesize,labelsep=period,format=plain,labelfont={bf}]{caption}

\usepackage{authblk}

\usepackage{hyperref}

\usepackage{enumitem}

\usepackage{tabu}

\renewcommand\rm[1]{\mathrm{#1}}

\newcommand\fn[1]{_{\mathrm{#1}}}
\newcommand\fnn[2]{_{\mathrm{#1},#2}}
\newcommand\fnx[1]{\fnn{#1}{x}}
\newcommand\fny[1]{\fnn{#1}{y}}

\newcommand\res{net} 
\newcommand\cm{cm} 

\newcommand\torque{\tau} 
\newcommand\moi{I} 

\newcommand\force{F} 
\newcommand\gforce{G} 
\newcommand\nforce{N} 
\newcommand\sforce{F} 

\newcommand\ex{\vec e_x}
\newcommand\ey{\vec e_y}
\newcommand\ez{\vec e_z}

\usepackage{tikz}
\usepackage{pgf}
\usepackage{pgfplots}
\pgfplotsset{compat=1.14}

\usetikzlibrary{3d,calc}
\usetikzlibrary{patterns,shapes}
\usetikzlibrary{arrows}
\usetikzlibrary{arrows.meta,bending,through,intersections,decorations.pathreplacing,decorations.markings}
\definecolor{tukblue}{RGB}{0,100,250}
\definecolor{tukbrightblue}{RGB}{0,132,212}
\definecolor{tukred}{RGB}{191,21,21}
\definecolor{tukbrightred}{RGB}{225,29,29}
\definecolor{tukgray}{RGB}{228,232,235}
\definecolor{tukmagenta}{RGB}{98,0,168}
\def\drawangle(#1)(#2)(#3){
  \draw[gray,shorten <=1pt,shorten >=1pt] ($(#1)!.3cm!(#2)$) -- ++ ($(#1)!.3cm!(#3)-(#1)$) -- ($(#1)!.3cm!(#3)$)
}

\def\cona{\rm A}
\def\conb{\rm B}
\def\con{\rm{A/B}}
\def\idtan{t}
\def\idtanvel{tangential velocity }

\begin{document}

\newcommand{\affilmark}[1]{\rlap{\textsuperscript{\itshape#1}}}

\twocolumn[{
\vspace*{3ex}
\begin{flushleft}\LARGE\bfseries\sffamily
Collision of two balls in a groove -- An interplay between translation and rotation
\end{flushleft}
\vspace*{2ex}
\begin{flushleft}\large
S Gröber and 
J Sniatecki\\[2ex]
\normalsize
Department of Physics, Technische Universität Kaiserslautern, Erwin-Schroedinger-Straße, D-67663 Kaiserslautern, Germany\\[1ex]
\upshape E-mail: \href{mailto:groeber@rhrk.uni-kl.de}{groeber@rhrk.uni-kl.de}
\end{flushleft}
\vspace*{3ex}

\centering
\begin{abstract}
\begin{minipage}{\dimexpr\textwidth-2cm}
\textbf{Abstract}~\\When two identical balls collide with equal inital speed on a horizontal groove, one can observe three trajectory types depending on the groove width. To explain this observation, we derive velocity diagrams of the balls motions from Newton’s laws of translation and rotation and kinematics of rigid bodies in a three-dimensional vectorial representation and compare them with experimental results. The velocity diagrams and an introduced determinant allow to discriminate between the trajectory types and to understand the interplay between translation and rotation after the collision of the balls.
\end{minipage}
\end{abstract}

\vspace*{3ex}
}]{}

\section{Introduction}

Over decades the motion of a single ball as well as the collision of balls influenced by impulsive, frictional or gravitational forces on flat or inclined surfaces have been analysed. Studies can be roughly categorized by their theoretical treatment of ball motions mainly with conservation principles (e.g. \cite{Domenech}, \cite{Redner}) or Newton's laws of translation and rotation (e.g. \cite{Hierrezuello}, \cite{Hopkins}) or a mixture of these principles and laws (e.g. \cite{Wallace}).

Keeping this in mind, we pick up an artificial laboratory phenomenon from \cite{Hanisch}, which can be performed with simple equipment as a qualitative freehand as well as a quantitative experiment (figure \ref{fig:experiment}): Two identical balls of uniform mass distribution, radius $R$ and mass $m$ are released from the same height on both inclined parts of a groove. The groove width $b\in(0,2R)$ is adjustable.

\begin{figure}
    \centering
    \includegraphics[width=8.2cm]{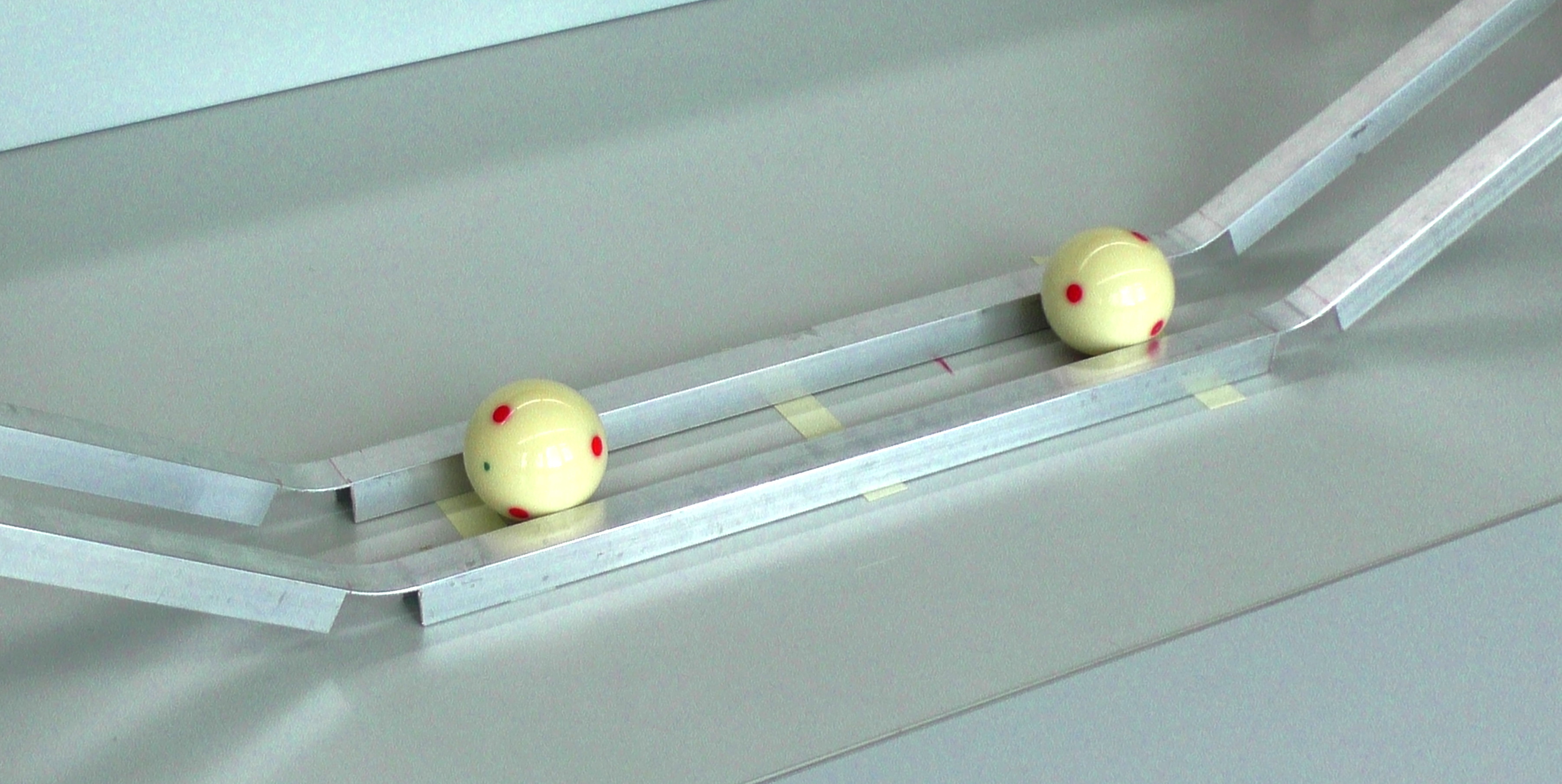}
    \caption{Setting of the experiment. The two inclined groove parts of the same inclination are used to generate equal initial ball speeds.}\label{fig:experiment}
\end{figure}
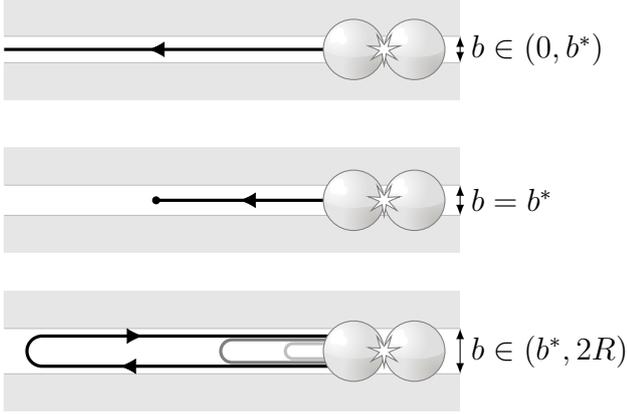
\begin{figure}
    \centering
    \begin{tikzpicture}
\def\vbb{.5}
\def\vlen{5}
\def\vxlen{1}

\def\drawgroove{
    \fill[gray!20] (-\vlen,\vbb+\vb*.5) rectangle (\vxlen,\vb*.5) coordinate (ur);
    \draw[gray!50] (-\vlen,\vb*.5) coordinate (ul) -- (\vxlen,\vb*.5);
    \fill[gray!20] (-\vlen,-\vbb-\vb*.5) rectangle (\vxlen,-\vb*.5);
    \draw[gray!50] (-\vlen,-\vb*.5) -- (\vxlen,-.5*\vb) coordinate(dr);
}

\def\vr{.4}

\begin{scope}[shift={(0,0)}]
    \def\vb{.35}
    \drawgroove

    \draw[very thick] (-\vr,0) -- (-\vlen,0);
    \foreach\x in {.6}{
        \draw[-{Latex[width=6pt,length=6pt]}] (-\vlen*\x,0) -- ++ (-.1,0);
    }

    \node[scale=\vr] at (-\vr,0) {\pgfimage{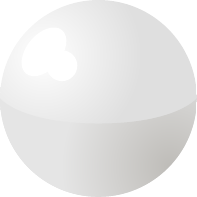}};
    \draw[gray] (-\vr,0) circle (\vr);
    \node[scale=\vr] at (\vr,0) {\pgfimage{ball.pdf}};
    \draw[gray] (\vr,0) circle (\vr);
    \node[gray,star,star points=7,draw,inner sep=.15cm,star point ratio=.3,fill=white] at (0,0) {};

    \draw[latex-latex] (ur) -- node[right]{$b\in(0,b^*)$} (dr);
\end{scope}

\begin{scope}[shift={(0,-2)}]
    \def\vb{.4}
    \drawgroove

    \draw[very thick] (-\vr,0) -- (-\vlen*.6,0);
    \foreach\x in {.6}{
        \draw[-{Latex[width=6pt,length=6pt]}] (-\vlen*\x*.6,0) -- ++ (-.1,0);
    }
    \fill (-\vlen*.6,0) circle (1.5pt);

    \node[scale=\vr] at (-\vr,0) {\pgfimage{ball.pdf}};
    \draw[gray] (-\vr,0) circle (\vr);
    \node[scale=\vr] at (\vr,0) {\pgfimage{ball.pdf}};
    \draw[gray] (\vr,0) circle (\vr);
    \node[gray,star,star points=7,draw,inner sep=.15cm,star point ratio=.3,fill=white] at (0,0) {};

    \draw[latex-latex] (ur) -- node[right]{$b=b^*$} (dr);
\end{scope}

\begin{scope}[shift={(0,-4)}]
    \def\vb{.6}
    \drawgroove

    \draw[very thick,gray] (-\vr,-.15) -- (-\vlen+3,-.15) arc (270:90:.15) -- (-\vr,.15);
    \draw[very thick,gray!50] (-\vr,-.10) -- (-\vlen+3.8,-.10) arc (270:90:.10) -- (-\vr,.10);
    \draw[very thick] (-\vr,-.2) -- (-\vlen+.5,-.2) arc (270:90:.2) -- (-\vr,.2);
    \foreach\x in {.75}{
        \draw[-{Latex[width=6pt,length=6pt]}] ({(-\vlen+.5)*\x},-.2) -- ++ (-.1,0);
        \draw[-{Latex[width=6pt,length=6pt]}] ({(-\vlen+.5)*\x+.1},.2) -- ++ (.1,0);
    }

    \node[scale=\vr] at (-\vr,0) {\pgfimage{ball.pdf}};
    \draw[gray] (-\vr,0) circle (\vr);
    \node[scale=\vr] at (\vr,0) {\pgfimage{ball.pdf}};
    \draw[gray] (\vr,0) circle (\vr);
    \node[gray,star,star points=7,draw,inner sep=.15cm,star point ratio=.3,fill=white] at (0,0) {};

    \draw[latex-latex] (ur) -- node[right]{$b\in(b^*,2R)$} (dr);
\end{scope}

\end{tikzpicture}
    \caption{Trajectory types of the left ball (radius $R$) after the first impact for different groove widths $b$ in respect to the critical groove width $b^*$.}\label{fig:cases}
\end{figure}

One can observe three trajectory types after the first impact of the balls on the horizontal part of the groove (figure \ref{fig:cases}):

\begin{itemize}[wide]
    \item If $b\in(0,b^*)$ is below a critical groove width $b^*$, the balls move away from each other.
    \item For $b=b^*$, the balls come to rest at a certain distance from each other.
    \item If $b\in(b^*,2R)$ is above the critical groove width $b^*$, the balls move away from each other, then move towards each other and collide again. This motion can appear several times with decreasing maximal distances of the balls.
\end{itemize}

The observation of these trajectory types is surprising in comparison with the seemingly similar experiment of carts colliding purely translational on an air track. This raises the question why the balls move in such a manner. The answer requires the understanding of the connection between causes (forces and torques) and effects (translational and rotational acceleration). Therefore, we perform here a theoretical treatment of the experiment with Newton's laws of translation and rotation and not with conservation principles \cite{Hanisch}, which consider only initial and finial states.

\section{Derivation of the balls motion}

\subsection{Assumptions, particle velocities and motion states}

In the following, we assume that the rolling and air friction forces acting on the balls are negligible and that the balls' collision is short, central and ideal-elastic. Moreover, we neglect tangential sliding forces between the balls during the impact. Before the first impact, the balls perform pure rolling with the same center-of-mass speed.

Generally, the velocity $\vec v\fn P$ of a particle in a point $\rm P$ of a ball in an inertial reference frame is
\begin{align}\begin{split}\vec v\fn P&=\vec v\fn\cm+\vec v\fn P'\\
&=\vec v\fn\cm+\vec\omega\times(\vec r\fn P-\vec r\fn\cm),\label{eq:VelocityField}\end{split}\end{align}
where $\vec v\fn\cm$ denotes the center-of-mass velocity in the inertial reference frame, $\vec v\fn P'$ the velocity of the particle $\rm P$ relative to the center of mass, $\vec\omega$ the angular velocity and $\vec r\fn P-\vec r\fn\cm$ the connecting vector from center of mass to $\rm P$.

\begin{figure}
    \centering
\def\drawball(#1:#2){
    \def\vrad{#1}
    \def\vangle{#2}
    \coordinate (cm) at (0,0);
    \coordinate (contact-right) at (-\vangle:\vrad);
    \coordinate (contact-left) at (180+\vangle:\vrad);
    \coordinate (outer-left) at ($(contact-left)+(-1,0)$);
    \coordinate (outer-right) at ($(contact-right)+(1,0)$);
    \coordinate (contact) at (0,{-sin{\vangle}*\vrad});

    \node[scale=\vrad] at (cm) {\pgfimage{ball.pdf}};
    \fill[gray!20] (outer-left) rectangle ($(outer-right)+(0,-.8)$);
    \draw[gray] (contact-right) arc (-\vangle:180+\vangle:\vrad);
    \draw[gray,densely dashed] (contact-right) arc (-\vangle:-180+\vangle:\vrad);
    \draw[gray] (outer-left) -- (outer-right);
}
\def\vlength{1.4}

\begin{tikzpicture}
\coordinate (origin) at (2.2,0);

\begin{scope}[shift={($-1*(origin)$)}]
\drawball(1.2:35)
\draw[tukblue,very thick,-latex] (contact) ++ (0,.05) -- ++ (-\vlength,0) node[above,pos=.5,font=\footnotesize]{$\vec v\fn\con$};
\draw[very thick,-latex,tukred] (contact) ++ (0,-.05) -- ++ (-\vlength,0) node[below,pos=.85,font=\footnotesize]{$\vec v\fn\con'$};
\draw[very thick,-latex,tukmagenta] (contact) -- node[below,pos=.7]{$\vec F\fn\con$} ++ (1.8,0);
\fill (cm) circle (2pt);
\fill (contact) circle (2pt) node[below,font=\footnotesize]{$\con$};;
\draw[very thick,-latex,tukred] (90+60:.6) arc (90+60:90-40:.6) node[pos=.5,above]{$\vec\omega$};
\node at (0,-2) {(a)};
\end{scope}

\begin{scope}[shift={(0,-3.4)}]
\drawball(1.2:35)
\draw[very thick,-latex,tukred] (contact) -- ++ (-\vlength,0) node[above,pos=.5,font=\footnotesize]{$\vec v\fn\con'$};
\draw[tukblue,very thick,-latex] (cm) -- ++ (\vlength,0) node[above,pos=.6,font=\footnotesize]{$\vec v\fn\cm$};
\fill(cm) circle (2pt);
\fill (contact) circle (2pt) node[below,font=\footnotesize]{$\con$};;
\draw[very thick,-latex,tukred] (90+60:.6) arc (90+60:90-40:.6) node[pos=.5,above]{$\vec\omega$};
\node at (0,-2) {(c)};
\end{scope}

\begin{scope}[shift={($1*(origin)$)}]
\drawball(1.2:35)
\draw[tukblue,very thick,-latex] (contact) -- ++ (-\vlength,0) node[above,pos=.5,font=\footnotesize]{$\vec v\fn\con$};
\draw[tukblue,very thick,-latex] (cm) -- ++ (-\vlength,0) node[above,pos=.5,font=\footnotesize]{$\vec v\fn\cm$};
\draw[very thick,-latex,tukmagenta] (contact) -- node[below,pos=.7]{$\vec F\fn\con$} ++ (1.8,0);
\fill (cm) circle (2pt);
\fill (contact) circle (2pt) node[below,font=\footnotesize]{$\con$};
\node at (0,-2) {(b)};
\end{scope}
\end{tikzpicture}
    \caption{Motion states pure slipping (a), pure gliding (b) and pure rolling (c) of a ball in a groove with center-of-mass velocity $\vec v\fn\cm$, velocity $\vec v\fn\con$ of particles in $\con$ with respect to an inertial reference frame, velocity $\vec v\fn\con'$ with respect to the center of mass, angular velocity $\vec\omega$ and kinetic friction forces $\vec F\fn\con$.}\label{fig:motion-states}
\end{figure}
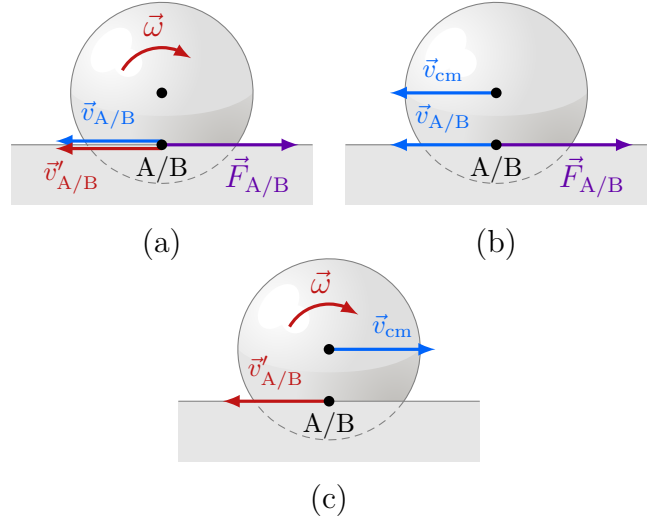

Especially the particles in $\cona$ and $\conb$, which are currently in contact with the groove and which have the identical velocity
\begin{align}\begin{split}\vec v\fn\con&=\vec v\fn\cm+\vec v\fn\con'\\
&=\vec v\fn\cm+\vec\omega\times(\vec r\fn\con-\vec r\fn\cm)\end{split}\label{eq:VelocityRigidBody}\end{align}
allow to categorize the balls motion states (figure \ref{fig:motion-states}): If one of the three velocities in \eqref{eq:VelocityRigidBody} is zero, the magnitudes of the other two velocities are equal and we get the motion states pure slipping (no translation), pure gliding (no rotation) and pure rolling. In all other cases we get intermediate motion states of rolling and gliding or rolling and slipping. If the ball does not perform pure rolling, i.e. the particles in $\rm A$ and $\rm B$ move relative to the groove ($\vec v\fn\con\neq\vec 0$), equal kinetic friction forces $\vec F\fn\con$ act on the balls in opposite direction to the velocity $\vec v\fn\con$.

\begin{figure}
    \centering
    \begin{tikzpicture}[]
\def\vgroovelen{7}
\def\vgroovewidth{3.2}
\def\vradius{1.8}
\def\vcorr{0};

\coordinate (z) at (-10:.707);
\coordinate (x) at (25:.707);

\begin{scope}[z={(z)},y={(0,1)},x={(x)}]
\coordinate (cm) at (-\vradius,{sqrt(\vradius^2-.25*\vgroovewidth^2)+\vcorr},0);
\coordinate (cm-other) at ($(cm)+(2*\vradius,0,0)$);
\coordinate (contact-left) at (-\vradius,0,\vgroovewidth*.5);
\coordinate (contact-right) at (-\vradius,0,-\vgroovewidth*.5);

\fill[gray!20] (-\vgroovelen*.5,0,-\vgroovewidth*.5) -- ++ (\vgroovelen,0,0) -- ++ (0,0,-1) -- ++ (-\vgroovelen,0,0) --cycle;
\fill[gray!40] (-\vgroovelen*.5,0,-\vgroovewidth*.5) -- ++ (\vgroovelen,0,0) -- ++ (0,-1,0) -- ++ (-\vgroovelen,0,0) --cycle;
\fill[gray!10] (-\vgroovelen*.5,0,-\vgroovewidth*.5) -- ++ (0,-1,0) -- ++ (0,0,-1) -- ++ (0,1,0) --cycle;
\end{scope}

\node[scale=\vradius] at (cm-other) {\pgfimage{ball.pdf}};
\node[scale=\vradius] at (cm) {\pgfimage{ball.pdf}};
\draw[gray] (cm) circle (\vradius);
\coordinate (leftball) at ($(cm)+(120:\vradius)$);

\begin{scope}[z={(z)},y={(0,1)},x={(x)}]
\fill[gray!20] (-\vgroovelen*.5,0,\vgroovewidth*.5) -- ++ (\vgroovelen,0,0) -- ++ (0,0,1) -- ++ (-\vgroovelen,0,0) --cycle;
\fill[gray!10] (-\vgroovelen*.5,0,\vgroovewidth*.5) -- ++ (0,-1,0) -- ++ (0,0,1) -- ++ (0,1,0) --cycle;
\fill[gray!40] (-\vgroovelen*.5,0,\vgroovewidth*.5+1) -- ++ (\vgroovelen,0,0) -- ++ (0,-1,0) -- ++ (-\vgroovelen,0,0) --cycle;

\draw[latex-latex,shorten >=2pt] (-\vradius,{sqrt(\vradius^2-.25*\vgroovewidth^2)},\vgroovewidth*.5) coordinate (rc) -- node[right]{$r$} (-\vradius,0,\vgroovewidth*.5);
\coordinate (circumference) at ($(rc)+(20:{sqrt(\vradius^2-.25*\vgroovewidth^2)})$);
\draw[gray,densely dashed] ($(cm)!1.5!(rc)$) -- ($(rc)!2.5!(cm)$);
\draw[gray] (rc) circle ({sqrt(\vradius^2-.25*\vgroovewidth^2)});
\draw[latex-latex,shorten <=2pt] (cm) -- node[left]{$R$} ++ (0,\vradius,0);

\draw[latex-latex,shorten <=2pt,shorten >=2pt] (contact-left) -- node[below]{$b$} (contact-right);

\draw[-latex] (cm) -- ++ (1,0,0) node[above,pos=.5]{$x$};
\draw[-latex] (cm) -- ++ (0,-.706,0) node[left,pos=.5]{$y$};
\draw[-latex] (cm) -- ++ (0,0,-1) node[above]{$z$};

\end{scope}

\draw[gray] (contact-left) -- ++ (-1.6,-1.2) coordinate (tip-contacts);
\draw[gray] (contact-right) -- (tip-contacts) node[below,font=\footnotesize,black]{contact points};
\draw[gray] (circumference) -- ++ (1,1) node[right,font=\footnotesize,black,text width=3cm]{circle of effective radius $r$};
\draw[gray] (leftball) -- ++ (120:.5) node[above,font=\footnotesize,black]{left ball};
\node[font=\footnotesize,black] at (1.8,-.8) {groove};

\fill (cm) circle (2pt) node[font=\footnotesize,below right=-1] {$\rm\cm$};
\fill (contact-left) circle (2pt) node[below right=-2]{$\cona$};
\fill (contact-right) circle (2pt) node[left]{$\conb$};

\end{tikzpicture}
\vspace*{-.5cm}
    \caption{Fixed coordinate system and groove width $b$, radius $R$ and effective radius $r$.}
    \label{fig:front-view}
\end{figure}
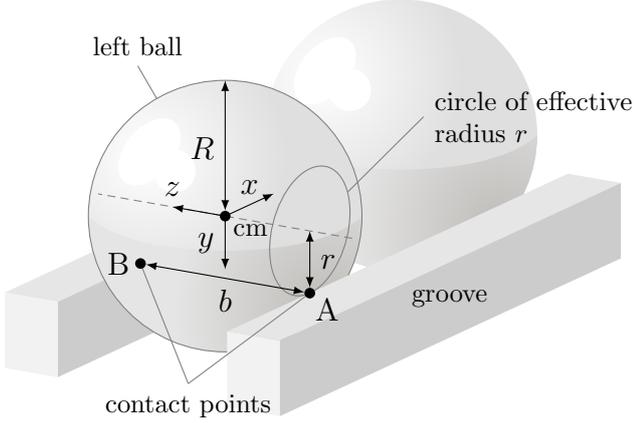

Especially for the fixed coordinate system in figure \ref{fig:front-view}, the connecting vectors are
\begin{align}\vec r\fn\con=r\ey\mp\tfrac b 2\ez\label{eq:ConnectingVector}\end{align}
(minus for $\cona$ and plus for $\conb$). Inserting \eqref{eq:ConnectingVector} in \eqref{eq:VelocityRigidBody} yields
\begin{align}\begin{split}\vec v\fn\con&=v\fnx\cm+(\omega_z\vec e_z)\times(r\ey\mp\tfrac b 2 \ez)\\
&=(v\fnx\cm-\omega_z r)\ex,\end{split}\label{eq:VelocityPointOfContact}\end{align}
In \eqref{eq:VelocityPointOfContact}, the tangential velocity $v\fn\idtan=\omega_z r$ of the particles $\rm A$ and $\rm B$ on the circle of effective radius $r$ can be positive or negative, depending on the sign of $\omega_z$ and the direction of rotation, respectively.

In the following, due to symmetry, we consider only the motion of the left ball. The origin of the coordinate system is at the position of the center of mass of the left ball at the time $t_0=0$, when the balls collide (figure \ref{fig:front-view}). Before the first impact ($t<t_0$) the ball performs pure rolling with center-of-mass velocity $v\fnx\cm>0$ and angular velocity $\omega_z>0$, i.e. $\vec v\fn\con=0$ and from \eqref{eq:VelocityPointOfContact} follows the rolling condition
\begin{align}v\fnx\cm=v\fn t.\end{align}
Due to the assumptions, right after the impact, the center-of-mass velocity changed only in direction whereas the angular velocity and the \idtanvel $v\fn\idtan$ keeps unchanged, i.e. \begin{align}-v\fnx\cm(t_0)=v\fn\idtan(t_0)\label{eq:AssumptionAfterImpact}.\end{align}
The rolling condition is no longer fulfilled and kinetic friction forces act on the ball.

\subsection{Center-of-mass and angular acceleration}\label{section:centerofmass-angularacceleration}

To determine the center-of-mass acceleration $\vec a\fn\cm$ and the angular acceleration $\vec\alpha$ of the ball after the first impact, we apply Newton's laws of translation and rotation
\begin{align}
    \vec\force\fn\res&=m\vec a\fn\cm\label{eq:NewtonTranslation}\\
    \text{and}\qquad\vec\torque\fn\res&=\moi\vec\alpha.\label{eq:NewtonRotation}
\end{align}
$\vec\force\fn\res$ denotes the net force on the ball, $\vec\torque\fn\res$ the net torque with respect to the center-of-mass and $\moi=\frac 5 2 m R^2$ the moment of inertia of the ball with respect to an axis passing through the center of mass.

For both laws, we need to identify all forces exerted on the ball. The groove exerts normal forces $\vec\nforce\fn\cona$ and $\vec\nforce\fn\conb$ on the ball in the contact points $\cona$ and $\conb$, respectively. Because the ball moves only in or against the $x$-direction, the $y$- and $z$-component of the net force is zero. Therefore, the normal forces and the weight force in the $y$-$z$-plane compensate each other, i.e. \begin{align}\vec\nforce\fn\cona+\vec\nforce\fn\conb=-mg\vec e_y.\end{align}

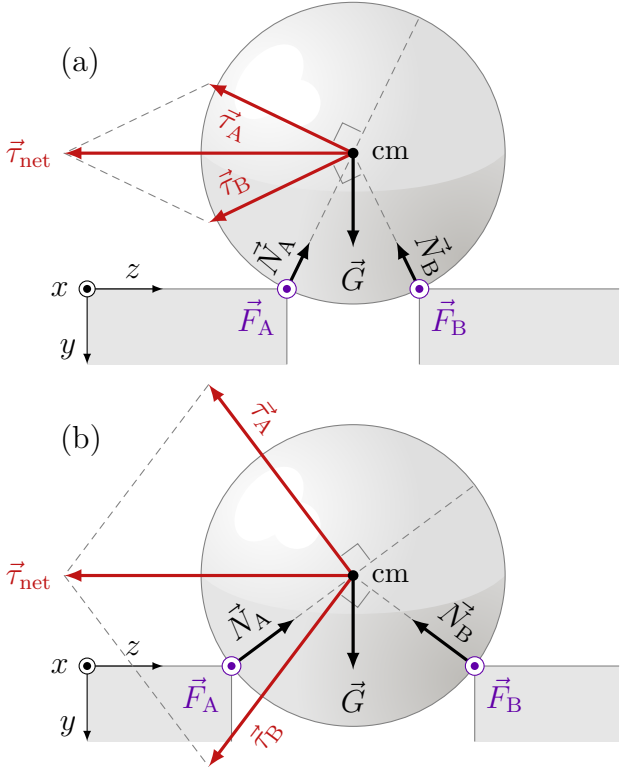
\begin{figure}[ht]
    \centering
    \begin{tikzpicture}

\coordinate (origin-1) at (0,0);
\coordinate (origin-2) at (0,-5);

\def\drawball[#1:#2]{
    \def\ballr{#1}
    \def\ballR{#2}
    \def\taufactor{1.9}
    \def\gforce{1.25}

    \coordinate (contact-left) at ({-sqrt(\ballR^2-\ballr^2)},0);
    \coordinate (contact-right) at ({sqrt(\ballR^2-\ballr^2)},0);
    \coordinate (cm) at (0,\ballr);

    \coordinate (outer-left) at (-3.5,0);
    \coordinate (outer-right) at (3.5,0);

    \node[scale=\ballR] at (cm) {\pgfimage{ball.pdf}};
    \draw[gray] (cm) circle (\ballR);

    \fill[gray!20] (outer-left) rectangle ($(contact-left)+(0,-1)$);
    \draw[gray] (outer-left) -- (contact-left) -- ++ (0,-1);
    \fill[gray!20] (outer-right) rectangle ($(contact-right)+(0,-1)$);
    \draw[gray] (outer-right) -- (contact-right) -- ++ (0,-1);

    \coordinate (fg) at ($(cm)+(0,-\gforce)$);
    \coordinate (na) at ($(contact-left)+({1/(2*\ballr)*\gforce*sqrt(\ballR^2-\ballr^2)},{1/2*\gforce})$);
    \coordinate (nb) at ($(contact-right)+({-1/(2*\ballr)*\gforce*sqrt(\ballR^2-\ballr^2)},{1/2*\gforce})$);
    \coordinate (taub) at ($(cm)+({-\ballR/2*\taufactor},{-\taufactor*\ballR*sqrt(\ballR^2-\ballr^2)/(2*\ballr)})$);
    \coordinate (taua) at ($(cm)+({-\ballR/2*\taufactor},{\taufactor*\ballR*sqrt(\ballR^2-\ballr^2)/(2*\ballr)})$);
    \coordinate (tau) at ($(cm)+({-2*\ballR/2*\taufactor},0)$);

    \draw[densely dashed,gray] (contact-left) -- ($(contact-left)!\ballR!(cm)$);
    \draw[densely dashed,gray] (contact-right) -- (cm);

    \begin{scope}[shift={(outer-left)}]
        \draw[-latex] (0,0) -- (1,0) node[above,pos=.6]{$z$};
        \draw[-latex] (0,0) -- (0,-1) node[left,pos=.8]{$y$};
        \fill[white,draw=black] (0,0) circle (3pt) node[left=.1cm,black]{$x$};
        \fill (0,0) circle (1.5pt);
    \end{scope}
}

\def\drawfrictions{
    \fill[white,draw=tukmagenta] (contact-left) circle (3.5pt) node[below left,tukmagenta]{$\vec F\fn\cona$};
    \fill[tukmagenta] (contact-left) circle (1.5pt);
    \fill[white,draw=tukmagenta] (contact-right) circle (3.5pt) node[below right,tukmagenta]{$\vec F\fn\conb$};
    \fill[tukmagenta] (contact-right) circle (1.5pt);
}

\begin{scope}[shift={(origin-1)}]
\node[anchor=west] at (-4,3) {(a)};
\drawball[1.8:2]

\drawangle(cm)($(contact-left)!1.2!(cm)$)(taua);
\drawangle(cm)(contact-right)(taub);

\draw[densely dashed,gray] (taua) -- (tau) -- (taub);
\draw[very thick,-latex,tukred] (cm) -- node[below,sloped,pos=.75]{$\vec\tau\fn\cona$} (taua);
\draw[very thick,-latex,tukred] (cm) -- node[above,sloped,pos=.75]{$\vec\tau\fn\conb$} (taub);
\draw[very thick,-latex] (contact-left) -- node[above,sloped]{$\vec\nforce\fn\cona$} (na);
\draw[very thick,-latex] (contact-right) -- node[above,sloped]{$\vec\nforce\fn\conb$} (nb);
\draw[very thick,-latex,tukred] (cm) -- (tau) node[left]{$\vec\tau\fn\res$};
\draw[very thick,-latex] (cm) -- (fg) node[below]{$\vec G$};

\drawfrictions
\fill (cm) circle (2pt) node[right=.1cm,font=\small] {$\rm\cm$};
\end{scope}

\begin{scope}[shift={(origin-2)}]
    \node[anchor=west] at (-4,3) {(b)};
\drawball[1.2:2]

\drawangle(cm)($(contact-left)!1.2!(cm)$)(taua);
\drawangle(cm)(contact-right)(taub);

\draw[densely dashed,gray] (taua) -- (tau) -- (taub);
\draw[very thick,-latex,tukred] (cm) -- node[above,sloped,pos=.75]{$\vec\tau\fn\cona$} (taua);
\draw[very thick,-latex,tukred] (cm) -- node[below,sloped,pos=.75]{$\vec\tau\fn\conb$} (taub);
\draw[very thick,-latex] (contact-left) -- node[above,sloped]{$\vec\nforce\fn\cona$} (na);
\draw[very thick,-latex] (contact-right) -- node[above,sloped]{$\vec\nforce\fn\conb$} (nb);
\draw[very thick,-latex,tukred] (cm) -- (tau) node[left]{$\vec\tau\fn\res$};
\draw[very thick,-latex] (cm) -- (fg) node[below]{$\vec G$};

\drawfrictions
\fill (cm) circle (2pt) node[right=.1cm,font=\small] {$\rm\cm$};
\end{scope}

\end{tikzpicture}
    \caption{Torques $\tau\fn\cona$, $\tau\fn\conb$ and $\tau\fn\res$ due to the kinetic friction forces $\vec F\fn\cona$ and $\vec F\fn\conb$, normal forces $\vec N\fn\cona$ and $\vec N\fn\conb$ and weight force $\vec\gforce$ for a small (a) and greater groove width $b$ (b). In (b) the magnitude of the kinetic friction forces is greater.}\label{fig:Torques}
\end{figure}

Due to symmetry, the $y$-components of the normal forces are
\begin{align}\nforce\fny\con=-\frac 1 2 m g.\label{eq:FrictionalForce}\end{align}
According to figure \ref{fig:Torques} we get
\begin{align}\frac{|\nforce\fny\con|}{|\vec\nforce\fn\con|}=\frac{r}{R}=k,\label{eq:FrictionalForceRatio}\end{align}
with the ratio
\begin{align}
k=\frac r R=\sqrt{1-\left(\frac b {2R}\right)^2},\label{eq:Ratio}
\end{align}
derived from figure \ref{fig:front-view}. From \eqref{eq:FrictionalForce} and \eqref{eq:FrictionalForceRatio} we get
\begin{align}|\vec\nforce\fn\con|=\frac 1{2k} m g.\end{align}
The normal forces produce kinetic friction forces $\vec F\fn\cona$ and $\vec F\fn\conb$ in the contact points $\cona$ and $\conb$, respectively. These forces point in $x$-direction, because they are opposed to the velocity
\begin{align}\begin{split}v\fnx\con(t_0)&=v\fnx\cm(t_0)-r\omega_z(t_0)\\&=2v\fnx\cm(t_0)<0\end{split}\end{align}
in $x$-direction. Application of the kinetic friction law yields
\begin{align}\vec\sforce\fn\con=\mu\fn k |\vec N\fn\con|\ex=\frac 1 {2k} \mu\fn k m g\ex,\label{eq:FrictionForce}\end{align}
where $\mu\fn k$ is the coefficient of kinetic friction.
Overall, the net force on the ball is
\begin{align}\begin{split}\vec F\fn\res&=\vec\sforce\fn\cona+\vec\sforce\fn\conb+\underset{=\vec 0}{\underbrace{\vec\nforce\fn\cona+\vec\nforce\fn\conb+\vec\gforce}}\\[-1em]&=\frac 1 k \mu\fn k m g\ex.\end{split}\label{eq:ResultantForce}\end{align}
According to \eqref{eq:NewtonTranslation} the center-of-mass's acceleration
\begin{align}\vec a\fn\cm=\frac 1 k \mu\fn k g\ex.\label{eq:LinearAcceleration}\end{align}
is constant, independent from the balls mass and dependent from the groove width.

For each force, we calculate the torque with respect to the center of mass. The lines of action of the normal forces $\vec\nforce\fn\con$ and the weight force $\vec\gforce$ pass through the center of mass, thus their torques are zero. The kinetic friction force $\vec\sforce\fn\cona$ and $\vec\sforce\fn\conb$ causes the torque $\vec\torque\fn\cona$ and $\vec\torque\fn\conb$, respectively. The torques are
\begin{align}\vec\torque\fn\con=(\vec r\fn\con-\vec r\fn\cm)\times\vec\sforce\fn\con.\label{eq:FrictionalTorque}\end{align}
Inserting \eqref{eq:ConnectingVector} and \eqref{eq:FrictionForce} in \eqref{eq:FrictionalTorque} yields
\begin{align}\begin{split}
\vec\torque\fn\con&=(r\ey\mp\tfrac b 2\ez)\times(\tfrac 1 {2k}\mu\fn k m g\ex)\\
&=-\tfrac 1{2k}\mu\fn k m g r\ez\mp \tfrac b{4k}\mu\fn k m g \ey.
\end{split}\end{align}
Hence, the $y$-component of the net torque $\vec\torque\fn\res=\vec\torque\fn\cona+\vec\torque\fn\conb$ is zero and we get
\begin{align}\begin{split}\vec\torque\fn\res&=-\frac 1 k \mu\fn k m g r\ez=(r\ey)\times(2\vec F\fn\con)\\&=-R\mu\fn k m g\ez.\label{eq:ResultantTorque}\end{split}\end{align}
The net torque $\tau\fn\res$ is independent of the groove width $b$, because for increasing $b$ the normal forces, the kinetic friction forces and thus their torques increase, but also the angle between the torques (figure \ref{fig:Torques}). 
According to \eqref{eq:NewtonRotation} and \eqref{eq:ResultantTorque} the angular acceleration
\begin{align}\vec\alpha=-\frac{5\mu\fn k g}{2R}\ez.\label{eq:RotationalAcceleration}\end{align}
is constant and independent from the balls mass and the groove width.

\subsection{Interplay between translation and rotation}

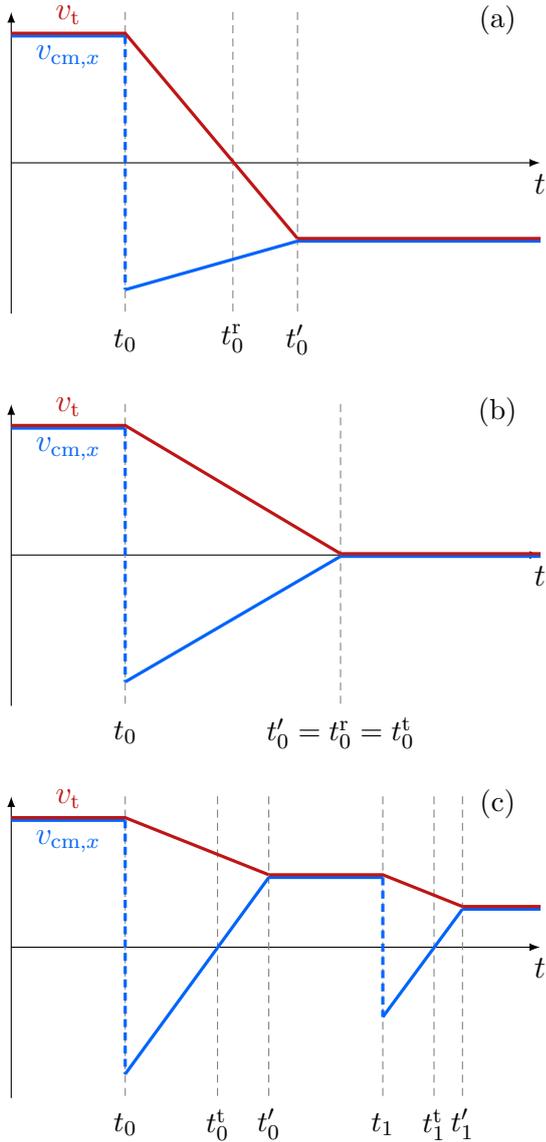
\begin{figure}[ht]
    \centering
    \begin{tikzpicture}

\def\tmax{\textwidth*0.4}
\def\vmax{2}
\def\vdist{1.2}

\def\voffset{.5}

\def\drawaxis{
    \draw[-latex] (0,0) -- (\tmax,0) node[below]{$t$};
    \draw[-latex] (0,-\vmax) -- (0,\vmax);
}
\foreach\i in {1,2,3}{
    \coordinate (origin-\i) at (0,{-(\vmax*2+\vdist)*\i});
}

\def\calcgraph[#1][#2,#3]{
    \clip (0,\vmax*1.2) rectangle (\tmax,-\vmax*1.3);
    \def\vdvcm{#2}
    \def\vdvt{#3}
    \coordinate (vcm-0-0) at (0,\vmax*.85);
    \coordinate (vt-0-0) at (0,\vmax*.85);
    \coordinate (c) at (vcm-0-0);

    \foreach\i/\deltat in {#1} {
        \coordinate (vcm-\i-1) at ($(c)+(\deltat,0)$);
        \path let \p1 = (vcm-\i-1) in (\x1,-\y1) coordinate (vcm-\i-1-reflected) (\x1,0) coordinate (t-\i-0);
        \coordinate (vt-\i-1) at ($(c)+(\deltat,0)$);
        \path[name path global/.expanded=vcm-slope-\i] (vcm-\i-1-reflected) -- ++ (10,{10*\vdvcm});
        \path[name path global/.expanded=vt-slope-\i] (vt-\i-1) -- ++ (10,{10*\vdvt});
        \path[name intersections={of={vcm-slope-\i} and {vt-slope-\i},by=c}];
        \coordinate (vcm-\i-2) at (c);
        \coordinate (vt-\i-2) at (c);
        \path let \p1 = (vcm-\i-2) in (\x1,0) coordinate (t-\i-2);
    }
}

\def\drawgraph[#1]{
    \coordinate (p-vcm) at ([yshift=-\voffset]vcm-0-0);
    \coordinate (p-vt) at ([yshift=\voffset]vt-0-0);

    \path (vcm-0-0) -- node[below,tukblue]{$v\fnx\cm$} (vcm-0-1);
    \path (vt-0-0) -- node[above,tukred]{$v\fn\idtan$} (vt-0-1);

    \foreach\i in {#1} {
        \draw[very thick,tukblue] (p-vcm) -- ([yshift=-\voffset]vcm-\i-1) coordinate (p-vcm);
        \draw[very thick,tukblue,densely dashed] (p-vcm) -- ([yshift=\voffset]vcm-\i-1-reflected) coordinate (p-vcm);
        \draw[very thick,tukblue] (p-vcm) -- ([yshift=-\voffset]vcm-\i-2) coordinate (p-vcm);
    
        \draw[very thick,tukred] (p-vt) -- ([yshift=\voffset]vt-\i-1) -- ([yshift=\voffset]vt-\i-2) coordinate (p-vt);
    }
    \draw[very thick,tukblue] (p-vcm) -- ++ (10,0);
    \draw[very thick,tukred] (p-vt) -- ++ (10,0);
}

\def\calczero(#1)[#2]{
    \path[name path global/.expanded=null] (0,0) -- (\tmax,0);
    \path[name intersections={of={#2-slope-#1} and {null},by=c}];
    \path let \p1 = (c) in (\x1,0) coordinate (t-#1-1);
}

\def\drawtime(#1)[#2]{
    \draw[gray,densely dashed] (t-#1) ++ (0,\vmax) -- ++ (0,-2*\vmax) node[below,black,font=\small] {#2};
}

\begin{scope}[shift={(origin-1)}]
\drawaxis
\calcgraph[0/1.5][.3,-1.2]
\calczero(0)[vt]

\drawtime(0-0)[$t_0\vphantom{t_0'}$]
\drawtime(0-1)[$t_0^{\rm r}$]
\drawtime(0-2)[$t_0'$]
\drawgraph[0]

\node[fill=white,inner sep=0,font=\small] at (6.4,1.9) {(a)};
\end{scope}

\begin{scope}[shift={(origin-2)}]
\drawaxis
\calcgraph[0/1.5][.6,-.6]

\drawtime(0-0)[$t_0\vphantom{t_0'}$]
\drawtime(0-2)[$t_0'=t_0^{\rm r}=t_0^{\rm t}$]
\drawgraph[0]

\node[fill=white,inner sep=0,font=\small] at (6.4,1.9) {(b)};
\end{scope}

\begin{scope}[shift={(origin-3)}]
\drawaxis
\calcgraph[0/1.5,1/1.5][1.4,-.4]
\calczero(0)[vcm]
\calczero(1)[vcm]

\drawtime(0-0)[$t_0\vphantom{t_0'}$]
\drawtime(0-1)[$t_0^{\rm t}$]
\drawtime(0-2)[$t_0'$]
\drawtime(1-0)[$t_1\vphantom{t_1'}$]
\drawtime(1-1)[$t_1^{\rm t}$]
\drawtime(1-2)[$t_1'$]
\drawgraph[0,1]

\node[fill=white,inner sep=0,font=\small] at (6.4,1.9) {(c)};
\end{scope}

\end{tikzpicture}
    \caption{Theoretical graphs of the center-of-mass' velocity $v\fnx\cm(t)$ and the tangential velocity $v\fn\idtan(t)$ for groove widths $b\in(0,b^*)$ (a), $b=b^*$ (b) and $b\in(b^*,2R)$ (c).}\label{fig:theoretical-graphs}
\end{figure}

Integrating \eqref{eq:LinearAcceleration} and \eqref{eq:RotationalAcceleration} over the time intervall $[t_0,t]$ yields with \eqref{eq:AssumptionAfterImpact} the linear functions
\begin{align}v\fnx\cm(t)&=v\fnx\cm(t_0)+\frac{\mu\fn k g}{k}t,\label{eq:VelocityCenterOfMass}\\
v\fn t(t)=\omega_z(t) r&=-v\fnx\cm(t_0)-\frac{5 k\mu\fn k g}{2}t.\label{eq:TangentialVelocity}\end{align}
The signs in \eqref{eq:VelocityCenterOfMass} and \eqref{eq:TangentialVelocity} indicate, that the graphs of $v\fnx\cm(t)$ and $v\fn\idtan(t)$ have for all $k$ an intersection point at time $t_0'$ (figure \ref{fig:theoretical-graphs}), when the ball fulfills the rolling condition after the first impact, i.e.
\begin{align}v\fnx\cm(t_0')=v\fn\idtan(t_0').\label{eq:RollingConditionAfterImpact}\end{align}
From \eqref{eq:VelocityCenterOfMass}-\eqref{eq:RollingConditionAfterImpact} we obtain for $t_0'$, $v\fnx\cm(t_0')$ and the position $x(t_0')=\int_0^{t_0'} v\fnx\cm(t)\,\mathrm dt$:
\begin{align}t_0'&=-\frac{4k}{2+5k^2}\frac{v\fnx\cm(t_0)}{\mu\fn k g}>0\\[.5em]
v\fnx\cm(t_0')&=-\frac{2-5k^2}{2+5 k^2}v\fnx\cm(t_0)=v\fn t(t_0')\label{eq:VelocityRollingCondition}\\[.5em]
x\fn\cm(t_0')&=-\frac{20 k^3}{2+5k^2}\frac{v\fnx\cm^2(t_0)}{\mu\fn k g}<0.\label{eq:PositionRollingCondition}\end{align}

\eqref{eq:VelocityRollingCondition} allows to discriminate between the three trajectory types:

\begin{itemize}[wide]
    \item $v\fnx\cm(t_0')<0$ for $b\in(0,b^*)$: The ball performs pure rolling for $t>t_0'$ against the $x$-direction and the balls move apart from each other. The center-of-mass velocity $v\fnx\cm$ remains negative after the impact, but the tangential velocity $v\fn t$ and the angular velocity $\vec\omega$ changes their direction according to \eqref{eq:TangentialVelocity} at the time
    \begin{align}t_0^{\rm r}=-\frac{2}{5k\mu\fn k g}v\fnx\cm(t_0)\in[t_0,t_0'].\end{align}
    At this time, the ball performs pure gliding (figure \ref{fig:motion-states}(b)).
    \item $v\fnx\cm(t_0')=0$ for $b=b^*$, i.e. the balls come to rest at time $t_0'$. From \eqref{eq:VelocityRollingCondition} and \eqref{eq:Ratio} we get for the critical groove width:
    \begin{align}\frac{b^*}{R}=\sqrt{\frac{12} 5}.\end{align}
    Neither the direction of translation nor the direction of rotation changes, because the rolling condition is fulfilled at the moment the center-of-mass velocity is zero.
    \item $v\fnx\cm(t_0')>0$ for $b\in(b^*,2R)$: The ball is rolling in $x$-direction for $t\geq t_0'$ and the balls collide again. The direction of rotation remains the same, whereas the direction of translation changes according to \eqref{eq:VelocityCenterOfMass} at the time
    \begin{align}t_0^{\rm t}=-\frac{k}{\mu\fn k g}v\fnx\cm(t_0)\in[t_0,t_0'].\end{align}
    At this time the ball performs pure slipping (figure \ref{fig:motion-states}(a)).
\end{itemize}

Overall, the slopes of the $v\fnx\cm$- and the $v\fn\idtan$-graph determine which of the three trajectory types occur. Therefore, the ratio
\begin{align}D=\frac{|a\fn t|}{|a\fnx\cm|}=\frac 5 2 k^2=\frac 5 2\left(1-\frac{b^2}{4R^2}\right)\end{align}
is a determinant for the three trajectory types, where
\begin{align}a\fn t=\frac{\rm d v\fn t}{\rm dt}=r\alpha_z\end{align}
denotes the tangential acceleration.

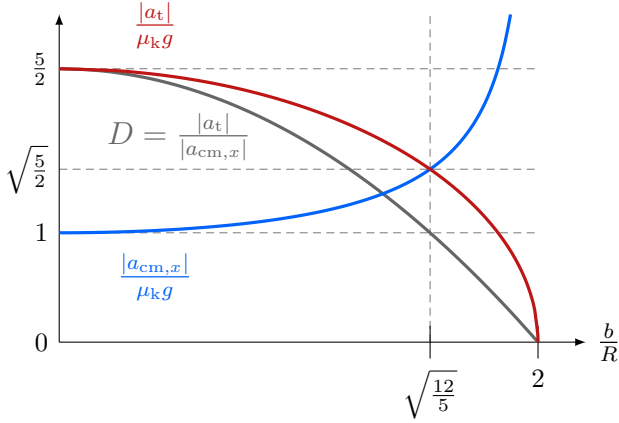
\begin{figure}
    \centering
    \begin{tikzpicture}

\def\vxwidth{6.3}
\def\vyheight{3}

\begin{scope}[xscale=\vxwidth,yscale=1.45]
\draw[gray,densely dashed] ({sqrt(3/5)},0) -- ++ (0,\vyheight);
\draw[gray,densely dashed] (0,1) node[left,black,font=\footnotesize]{$1$} -- ++ (1,0);
\draw[gray,densely dashed] (0,{5/2}) node[left,black,font=\footnotesize]{$\tfrac{5}{2}$} -- ++ (1,0);
\draw[gray,densely dashed] (0,{sqrt(5/2)}) node[left,black,font=\footnotesize]{$\sqrt{\tfrac 5 2}$} -- ++ (1,0);

\draw ({sqrt(3/5)},.15) -- ++ (0,-.3) node[below,font=\footnotesize]{$\sqrt{\frac{12} 5}$};
\draw (1,.15) -- ++ (0,-.3) node[below,font=\footnotesize]{$2$};

\draw[-latex] (0,0) node[left,font=\footnotesize]{$0$} -- (1.1,0) node[right]{$\frac b R$};
\draw[-latex] (0,0) -- (0,\vyheight);

\draw[very thick,black!60,smooth] plot[domain=0:1] (\x,{5/2*(1-\x^2)});
\draw[very thick,tukblue,smooth] plot[samples=80,domain=0:.943,smooth,tension=.5] (\x,{1/sqrt(1-\x^2)});
\draw[very thick,tukred,smooth] plot[samples=80,domain=0:1,smooth,tension=.5] ({sqrt(\x)},{5/2*sqrt(1-\x)}) -- (1,0);

\node[tukblue] at (.2,.6) {$\frac{|a\fnx\cm|}{\mu\fn k g}$};
\node[tukred] at (.2,2.9) {$\frac{|a\fn t|}{\mu\fn k g}$};
\node[black!60,inner sep=0] at (.25,1.88) {$D=\frac{|a\fn t|}{|a\fnx\cm|}$};
\end{scope}

\end{tikzpicture}
    \caption{Normalized center-of-mass acceleration $|a\fnx\cm|/(\mu\fn k g)$, normalized tangential acceleration $|a\fn t|/(\mu\fn k g)$ and ratio $D$ against the ratio $b/R$.}\label{fig:Determinant}
\end{figure}

For $D\in(1,\tfrac 5 2)$ and $b\in(0,b^*)$, the tangential acceleration $a\fn t$ is greater than the center-of-mass acceleration $a\fnx\cm$ (figure \ref{fig:Determinant}). Translatory motion decelerates faster than rotatory motion, and therefore, the direction of translation changes before the direction of rotation could change and the balls move apart. For $D\in(0,1)$ and $b\in(b^*,2R)$ the roles of translation and rotation are reversed, the balls move towards each other and collide several times. For $D=1$ and $b=b^*$ the center-of-mass and the tangential acceleration are equal in magnitude, i.e. the translatory and rotatory motion synchronously decelerate till the balls are at rest.

\begin{figure}
    \centering
    \begin{tikzpicture}
\pgfplotsset{
  /pgfplots/xlabel near ticks/.style={
     /pgfplots/every axis x label/.style={
        at={(ticklabel cs:0.5)},anchor=near ticklabel,font=\small
     }
  },
  /pgfplots/ylabel near ticks/.style={
    /pgfplots/every axis y label/.style={
     at={(ticklabel cs:0.5)},rotate=90,anchor=near ticklabel,font=\small}
  }
}
\pgfplotsset{every tick label/.append style={font=\footnotesize}}

\def\dy{5}

\begin{scope}[shift={(0,\dy)}]
\begin{axis}[
    /pgf/number format/.cd,/pgf/number format/fixed,
        1000 sep={},
axis y line=left,axis x line=bottom,
ylabel=Velocities in $\rm{m}/\rm{s}$,
clip=false,
y label style={at={(axis description cs:-0.1,.5)},anchor=south},
xlabel near ticks,
width=8.5cm,height=6cm,
xmin=-0.18,xmax=0.64,
ymin=-0.7,ymax=0.7,
legend style={at={(6.8cm,4.3cm)},anchor=north east},
grid=both,
minor x tick num=4,
minor y tick num=4,
minor grid style={very thin,gray!20},
]
\draw (axis cs:\pgfkeysvalueof{/pgfplots/xmin},0) -- (axis cs:\pgfkeysvalueof{/pgfplots/xmax},0);
\draw[densely dashed] (axis cs:0.18,\pgfkeysvalueof{/pgfplots/ymin}) -- (axis cs:0.18,\pgfkeysvalueof{/pgfplots/ymax}) node[pos=.1,left]{$t_0^{\rm r}$};
\draw[densely dashed] (axis cs:0.22,\pgfkeysvalueof{/pgfplots/ymin}) -- (axis cs:0.22,\pgfkeysvalueof{/pgfplots/ymax}) node[pos=.1,right]{$t_0'$};
\addplot[
    only marks,
    mark size=1pt,
    tukred
] table {
    -0.18	0.553630066
    -0.16	0.564944688
    -0.14	0.580263949
    -0.12	0.596306339
    -0.1	0.585533944
    -0.08	0.562538888
    -0.06	0.545641824
    -0.04	0.550011364
    -0.02	0.555005631
    0	0.527754678
    0.02	0.475532596
    0.04	0.435674764
    0.06	0.37098686
    0.08	0.303160436
    0.1	0.263082687
    0.12	0.198777514
    0.14	0.132216678
    0.16	0.088670739
    0.18	0.025124689
    0.2	-0.040697051
    0.22	-0.086204698
    0.24	-0.116216393
    0.26	-0.112805142
    0.28	-0.106066017
    0.3	-0.120104121
    0.32	-0.117739118
    0.34	-0.108915105
    0.36	-0.11643131
    0.38	-0.119399539
    0.4	-0.112971235
    0.42	-0.110255385
    0.44	-0.115244306
    0.46	-0.113192314
    0.48	-0.111495516
    0.5	-0.1125
    0.52	-0.110255385
    0.54	-0.106360237
    0.56	-0.106124926
    0.58	-0.106448344
    0.6	-0.112527774
    0.62	-0.110255385
    0.64	-0.1075    
};

\addplot[
    only marks,
    mark size=1pt,
    tukblue
] table{
    -0.18	0.61
    -0.16	0.6125
    -0.14	0.615
    -0.12	0.6075
    -0.1	0.605
    -0.08	0.605
    -0.06	0.5975
    -0.04	0.61
    -0.02	0.6125
    0	0.055
    0.02	-0.51
    0.04	-0.505
    0.06	-0.4525
    0.08	-0.4025
    0.1	-0.3675
    0.12	-0.325
    0.14	-0.2825
    0.16	-0.24
    0.18	-0.195
    0.2	-0.155
    0.22	-0.125
    0.24	-0.1125
    0.26	-0.1125
    0.28	-0.115
    0.3	-0.11
    0.32	-0.115
    0.34	-0.12
    0.36	-0.1125
    0.38	-0.1125
    0.4	-0.115
    0.42	-0.1125
    0.44	-0.1125
    0.46	-0.12
    0.48	-0.11
    0.5	-0.1075
    0.52	-0.11
    0.54	-0.11
    0.56	-0.115
    0.58	-0.115
    0.6	-0.1175
    0.62	-0.11
    0.64	-0.1075       
};
\legend{$v\fn\idtan(t)$,$v\fnx\cm(t)$}
\end{axis}
\end{scope}

\begin{scope}[shift={(0,0)}]
\begin{axis}[
    /pgf/number format/.cd,/pgf/number format/fixed,
        1000 sep={},
axis y line=left,axis x line=bottom,
ylabel=Velocities in $\rm{m}/\rm{s}$,
clip=false,
y label style={at={(axis description cs:-0.1,.5)},anchor=south},
xlabel near ticks,
width=8.5cm,height=6cm,
xmin=-0.2,xmax=0.44,
ymin=-0.7,ymax=0.7,
legend style={at={(6cm,4.5cm)},anchor=north east},
grid=both,
minor x tick num=4,
minor y tick num=4,
minor grid style={very thin,gray!20},
]
\draw (axis cs:\pgfkeysvalueof{/pgfplots/xmin},0) -- (axis cs:\pgfkeysvalueof{/pgfplots/xmax},0);
\draw[densely dashed] (axis cs:0.19,\pgfkeysvalueof{/pgfplots/ymin}) -- (axis cs:0.19,\pgfkeysvalueof{/pgfplots/ymax}) node[pos=.1,right]{$t_0'$};
\addplot[
    only marks,
    mark size=1pt,
    tukblue
] table{
    -0.2	0.555
    -0.18	0.555
    -0.16	0.545
    -0.14	0.545
    -0.12	0.5475
    -0.1	0.5425
    -0.08	0.545
    -0.06	0.5425
    -0.04	0.5475
    -0.02	0.3925
    0	-0.13
    0.02	-0.4625
    0.04	-0.4075
    0.06	-0.3575
    0.08	-0.3125
    0.1	-0.25
    0.12	-0.1875
    0.14	-0.14
    0.16	-0.095
    0.18	-0.055
    0.2	-0.0175
    0.22	0
    0.24	0
    0.26	0
    0.28	0
    0.3	0
    0.32	0
    0.34	0
    0.36	0
    0.38	0
    0.4	0
    0.42	0
    0.44	0    
};
\addplot[
    only marks,
    mark size=1pt,
    tukred
] table {
    -0.2	0.549010473
    -0.18	0.540231432
    -0.16	0.507352195
    -0.14	0.508066187
    -0.12	0.516532913
    -0.1	0.513255541
    -0.08	0.50000625
    -0.06	0.500312402
    -0.04	0.513578134
    -0.02	0.550420294
    0	0.517034815
    0.02	0.460061137
    0.04	0.430464865
    0.06	0.376065154
    0.08	0.308636113
    0.1	0.236603571
    0.12	0.15542281
    0.14	0.115027171
    0.16	0.078102497
    0.18	0.036400549
    0.2	-0.0125
    0.22	0
    0.24	0
    0.26	0
    0.28	0
    0.3	0
    0.32	0
    0.34	0
    0.36	0
    0.38	0
    0.4	0
    0.42	0
    0.44	0               
};
\end{axis}
\end{scope}

\begin{scope}[shift={(0,-\dy)}]
\begin{axis}[
    /pgf/number format/.cd,/pgf/number format/fixed,
        1000 sep={},
axis y line=left,axis x line=bottom,
xlabel=Time $t$ in $\rm s$,
ylabel=Velocities in $\rm{m}/\rm{s}$,
clip=false,
y label style={at={(axis description cs:-0.1,.5)},anchor=south},
xlabel near ticks,
width=8.5cm,height=6cm,
xmin=-0.18,xmax=0.88,
ymin=-0.9,ymax=0.9,
legend style={at={(6cm,4.5cm)},anchor=north east},
grid=both,
minor x tick num=4,
minor y tick num=4,
minor grid style={very thin,gray!20},
]
\draw (axis cs:\pgfkeysvalueof{/pgfplots/xmin},0) -- (axis cs:\pgfkeysvalueof{/pgfplots/xmax},0);
\draw[densely dashed] (axis cs:0.24,\pgfkeysvalueof{/pgfplots/ymin}) -- (axis cs:0.24,\pgfkeysvalueof{/pgfplots/ymax}) node[pos=.1,left]{$t_0^{\rm t}$};
\draw[densely dashed] (axis cs:0.29,\pgfkeysvalueof{/pgfplots/ymin}) -- (axis cs:0.29,\pgfkeysvalueof{/pgfplots/ymax}) node[pos=.1,right]{$t_0'$};
\addplot[
    only marks,
    mark size=1pt,
    tukblue
] table{
    -0.18	0.795
    -0.16	0.8
    -0.14	0.7875
    -0.12	0.79
    -0.1	0.795
    -0.08	0.7925
    -0.06	0.795
    -0.04	0.79
    -0.02	0.79
    0	0.0925
    0.02	-0.6425
    0.04	-0.635
    0.06	-0.5575
    0.08	-0.4825
    0.1	-0.41
    0.12	-0.3375
    0.14	-0.27
    0.16	-0.21
    0.18	-0.14
    0.2	-0.08
    0.22	-0.0275
    0.24	0.0325
    0.26	0.0975
    0.28	0.1525
    0.3	0.1825
    0.32	0.1925
    0.34	0.19
    0.36	0.195
    0.38	0.1925
    0.4	0.1825
    0.42	0.19
    0.44	0.195
    0.46	0.195
    0.48	0.1925
    0.5	0.19
    0.52	0.195
    0.54	0.19
    0.56	0.1875
    0.58	0.1925
    0.6	0.195
    0.62	0.195
    0.64	0.19
    0.66	0.1875
    0.68	0.1275
    0.7	-0.02
    0.72	-0.1075
    0.74	-0.0575
    0.76	0.005
    0.78	0.0375
    0.8	0.0475
    0.82	0.045
    0.84	0.04
    0.86	0.04
    0.88	0.03
};

\addplot[
    only marks,
    mark size=1pt,
    tukred
] table {
    -0.18	0.772537374
    -0.16	0.780346664
    -0.14	0.778118294
    -0.12	0.778708332
    -0.1	0.803009963
    -0.08	0.770848331
    -0.06	0.759494075
    -0.04	0.78094956
    -0.02	0.737843141
    0	0.781253039
    0.02	0.69923619
    0.04	0.687663617
    0.06	0.607150105
    0.08	0.567208075
    0.1	0.524171966
    0.12	0.494785307
    0.14	0.47463802
    0.16	0.406970515
    0.18	0.384398881
    0.2	0.355844067
    0.22	0.305951385
    0.24	0.277680122
    0.26	0.210059515
    0.28	0.16735068
    0.3	0.174427635
    0.32	0.182157212
    0.34	0.172789612
    0.36	0.187366619
    0.38	0.187766477
    0.4	0.188215302
    0.42	0.197230829
    0.44	0.189225395
    0.46	0.16988967
    0.48	0.19540023
    0.5	0.215014534
    0.52	0.181331878
    0.54	0.193471574
    0.56	0.199828051
    0.58	0.184814096
    0.6	0.173439471
    0.62	0.182517122
    0.64	0.198887531
    0.66	0.18179659
    0.68	0.182773767
    0.7	0.19139292
    0.72	0.13362728
    0.74	0.076485293
    0.76	0.0525
    0.78	0.0425
    0.8	0.050062461
    0.82	0.049117207
    0.84	0.040388736
    0.86	0.04472136
    0.88	0.05153882     
};
\end{axis}
\end{scope}

\end{tikzpicture}
    \caption{Measured center-of-mass velocity $v\fnx\cm(t)$ and tangential velocity $v\fn t(t)$ for ball radius $R=2.75\,\rm{cm}$, ball mass $m=0.167\,\rm{kg}$ and groove widths $b=3.5\,\rm{cm}$ (a), $b=4.4\,\rm{cm}$ (b) and $b=4.9\,\rm{cm}$ (c). Dashed lines indicate the calculated times for pure gliding (a), for pure slipping (c) and the beginning of pure rolling (a)-(c).}
\end{figure}
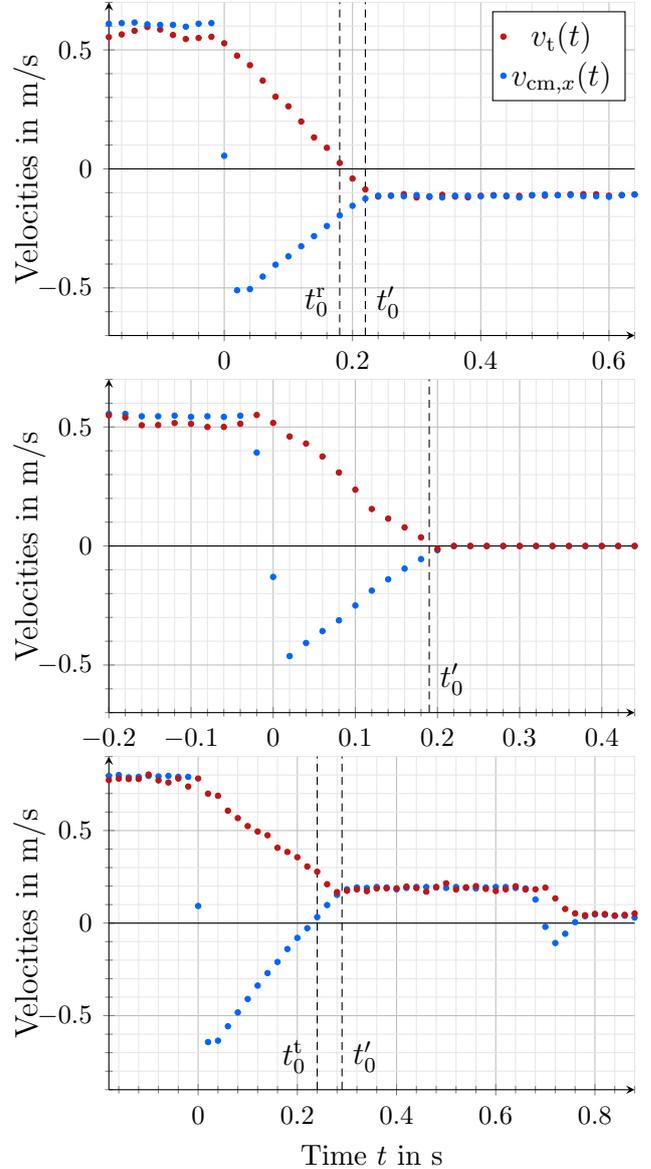

\section{Comparison of theory and experiment}
For groove widths $b$ of the three corresponding trajectory types, the balls translatory and rotatory motion was measured with video analysis \cite{Allain} by tracking the center of mass coordinates and the coordinates of a point on an circle of effective radius $r$ with respect to a fixed coordinate system (frame rate of 100/s). Coordinate measurement errors were minimized by sufficient high ball radius and a selection of smallest possible image section, which is adapted to the ball's motion. We checked the equality of the initial ball speeds and calculated the left ball's center-of-mass velocity $v\fnx\cm$ and according to \eqref{eq:VelocityPointOfContact} the tangential velocity $v\fn t$ (figure \ref{fig:theoretical-graphs}).

Equal magnitudes of $v\fnx\cm$ right before and after the collision as well as the constant $v\fnx\cm$ in pure rolling time intervalls confirm the previously made assumptions (see section \ref{section:centerofmass-angularacceleration}). The approximately linear graphs of the intermediate motion states confirm the use of the velocity-independent kinetic friction law \eqref{eq:FrictionForce}.

The experimental results in table \ref{table:results} indicate that in accordance with the theory (figure \ref{fig:Determinant}, \eqref{eq:ResultantForce} and \eqref{eq:ResultantTorque}) for increasing groove width $b$ the center-of-mass acceleration $a\fnx\cm$ and the net force $\vec F\fn\res$ are increasing and the tangential acceleration $a\fn t$ is decreasing, whereas the torque $\vec\tau\fn\res$ is approximately constant. The mean kinetic friction coefficient from table \ref{table:results} is $\mu\fn k=0.17$.

\begin{table}
    \tabulinesep=1.5mm
    \begin{tabu}{X[c1]|X[c1]|X[c1]|X[c1]}
        $b$ in cm & $|a\fnx\cm|$ in $\rm m/\rm s^2$ & $\mu\fn{k,cm}$ & $|\vec F\fn\res|$ in N\\\hline
        $3.5$ & $2.15$ & $0.17$ & $0.36$\\
        $4.4$ & $2.86$ & $0.18$ & $0.48$\\
        $4.9$ & $3.26$ & $0.17$ & $0.54$\\\hline\hline
        $b$ in cm & $|a\fn t|$ in $\rm m/\rm s^2$ & $\mu\fn{k,t}$ & $|\vec\tau\fn\res|$ in Nm\\\hline
        $3.5$ & $2.92$ & $0.15$ & $0.008$\\
        $4.4$ & $2.61$ & $0.17$ & $0.009$\\
        $4.9$ & $2.00$ & $0.16$ & $0.008$
    \end{tabu}
    \caption{Experimentally determined quantities for the intermediate motion states after the first collision (figure \ref{fig:experiment}). The center-of-mass acceleration $|a\fnx\cm|$ and the tangential acceleration $|a\fn t|$ is determined from linear regression of $v\fnx\cm(t)$ and $v\fn t(t)$. Corresponding kinetic friction coefficient $\mu\fn{k,cm}$ and $\mu\fn{k,t}$ is calculated from the slope in \eqref{eq:VelocityCenterOfMass} and \eqref{eq:TangentialVelocity}, respectively. The net force $|\vec F\fn\res|$ and the net torque $|\vec\tau\fn\res|$ are calculated from \eqref{eq:ResultantForce} and \eqref{eq:ResultantTorque} using $\mu\fn k=\mu\fn{k,cm}$.} 
    \label{table:results}
\end{table}

\section{Summary and educational conclusions}

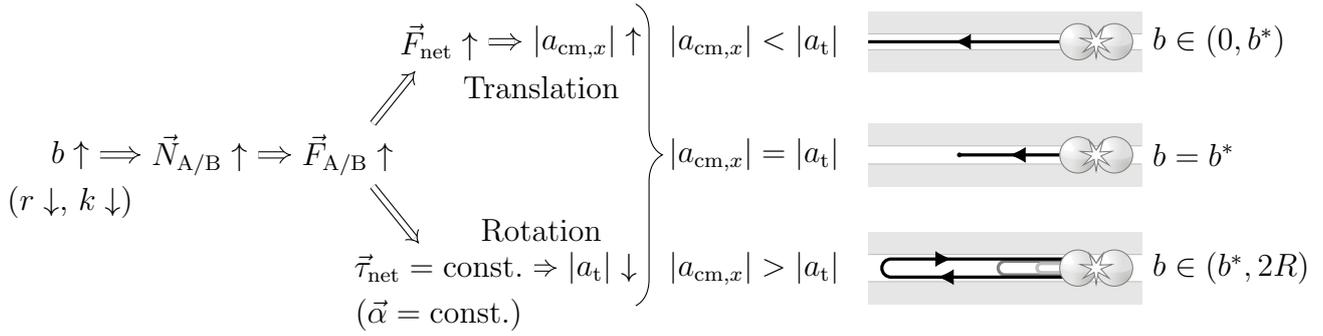
\begin{figure*}
    \centering
    \begin{tikzpicture}[node distance=2.2cm and 2cm,
    implies/.style={double,double equal sign distance,-implies}]

\def\vbb{.5}
\def\vlen{5}
\def\vxlen{1}

\def\drawgroove{
    \fill[gray!20] (-\vlen,\vbb+\vb*.5) rectangle (\vxlen,\vb*.5) coordinate (ur);
    \draw[gray!50] (-\vlen,\vb*.5) coordinate (ul) -- (\vxlen,\vb*.5);
    \fill[gray!20] (-\vlen,-\vbb-\vb*.5) rectangle (\vxlen,-\vb*.5);
    \draw[gray!50] (-\vlen,-\vb*.5) -- (\vxlen,-.5*\vb) coordinate(dr);
}

\def\vr{.4}

\node (groove-width) at (0,0) {$b\uparrow$};
\node[below of=groove-width,node distance=1.5em] {($r\downarrow$, $k\downarrow$)};

\node (normal-force) [right of=groove-width,shift={(-.5,0)}] {$\vec N\fn\con\uparrow$};
\node (friction-force) [right of=normal-force,shift={(-.25,0)}] {$\vec F\fn\con\uparrow$};
\node (f-res) [right of=friction-force,shift={(-1,1.5)}] {$\vec F\fn\res\uparrow$};
\node (t-res) [right of=friction-force,shift={(-1,-1.5)}] {$\vec\tau\fn\res=\mathrm{const.}$};
\node[below of=t-res,node distance=1.5em] {($\vec\alpha=\mathrm{const.}$)};
\node (at) [right of=t-res,shift={(-.05,0)},align=right] {$|a\fn t|\downarrow$};
\node (acm) [right of=f-res,shift={(-.25,0)},align=right] {$|a\fnx\cm|\uparrow$};

\draw[implies] (groove-width) -- (normal-force);
\draw[implies] (normal-force) -- (friction-force);
\draw[implies] (friction-force) -- (f-res);
\draw[implies] (friction-force) -- (t-res);
\draw[implies] (f-res) -- (acm);
\draw[implies] (t-res) -- (at);

\draw [decorate,decoration={brace,amplitude=10pt,raise=4pt}] ($(acm)+(.5,.5)$) -- ($(at)+(.3,-.5)$);
\node[] at (6.2,-1){Rotation};
\node[] at (6.2,.9){Translation};

\begin{scope}[scale=.6,shift={(22.5,0)}]
    \begin{scope}[shift={(0,2.5)}]
        \def\vb{.35}
        \drawgroove
    
        \draw[very thick] (-\vr,0) -- (-\vlen,0);
        \foreach\x in {.6}{
            \draw[-{Latex[width=6pt,length=6pt]}] (-\vlen*\x,0) -- ++ (-.1,0);
        }
    
        \node[scale=\vr*.6] at (-\vr,0) {\pgfimage{ball.pdf}};
        \draw[gray] (-\vr,0) circle (\vr);
        \node[scale=\vr*.6] at (\vr,0) {\pgfimage{ball.pdf}};
        \draw[gray] (\vr,0) circle (\vr);
        \node[gray,star,star points=7,draw,inner sep=.15cm,star point ratio=.3,fill=white] at (0,0) {};
    
        \path (ur) -- node[right]{$b\in(0,b^*)$} (dr);

        \path[] (-9,0) -- node[]{$|a\fnx\cm|<|a\fn t|$} (-6,0);
    \end{scope}
    
    \begin{scope}[shift={(0,0)}]
        \def\vb{.4}
        \drawgroove
    
        \draw[very thick] (-\vr,0) -- (-\vlen*.6,0);
        \foreach\x in {.6}{
            \draw[-{Latex[width=6pt,length=6pt]}] (-\vlen*\x*.6,0) -- ++ (-.1,0);
        }
        \fill (-\vlen*.6,0) circle (1.5pt);
    
        \node[scale=\vr*.6] at (-\vr,0) {\pgfimage{ball.pdf}};
        \draw[gray] (-\vr,0) circle (\vr);
        \node[scale=\vr*.6] at (\vr,0) {\pgfimage{ball.pdf}};
        \draw[gray] (\vr,0) circle (\vr);
        \node[gray,star,star points=7,draw,inner sep=.15cm,star point ratio=.3,fill=white] at (0,0) {};
    
        \path (ur) -- node[right]{$b=b^*$} (dr);
        \path[] (-9,0) -- node[]{$|a\fnx\cm|=|a\fn t|$} (-6,0);
    \end{scope}
    
    \begin{scope}[shift={(0,-2.5)}]
        \def\vb{.6}
        \drawgroove
    
        \draw[very thick,gray] (-\vr,-.15) -- (-\vlen+3,-.15) arc (270:90:.15) -- (-\vr,.15);
        \draw[very thick,gray!50] (-\vr,-.10) -- (-\vlen+3.8,-.10) arc (270:90:.10) -- (-\vr,.10);
        \draw[very thick] (-\vr,-.2) -- (-\vlen+.5,-.2) arc (270:90:.2) -- (-\vr,.2);
        \foreach\x in {.75}{
            \draw[-{Latex[width=6pt,length=6pt]}] ({(-\vlen+.5)*\x},-.2) -- ++ (-.1,0);
            \draw[-{Latex[width=6pt,length=6pt]}] ({(-\vlen+.5)*\x+.1},.2) -- ++ (.1,0);
        }
    
        \node[scale=\vr*.6] at (-\vr,0) {\pgfimage{ball.pdf}};
        \draw[gray] (-\vr,0) circle (\vr);
        \node[scale=\vr*.6] at (\vr,0) {\pgfimage{ball.pdf}};
        \draw[gray] (\vr,0) circle (\vr);
        \node[gray,star,star points=7,draw,inner sep=.15cm,star point ratio=.3,fill=white] at (0,0) {};
    
        \path (ur) -- node[right]{$b\in(b^*,2R)$} (dr);
        \path[] (-9,0) -- node[]{$|a\fnx\cm|>|a\fn t|$} (-6,0);
    \end{scope}
\end{scope}
\end{tikzpicture}
    \caption{Cause-and-effect chain to explain the occurence of the three trajectory types with groove width $b$, effective radius $r$, ratio $k=r/R$, normal forces $\vec N\fn\con$, kinetic friction forces $\vec F\fn\con$, net force $\vec F\fn\res$, net torque $\vec\tau\fn\res$, angular acceleration $\vec\alpha$, center-of-mass acceleration $a\fnx\cm$ and tangential acceleration $a\fn t$. $\uparrow$ and $\downarrow$ denotes increasing and decreasing in magnitude, respectively.}\label{fig:chain}
\end{figure*}

In this article, we derived a cause-and-effect chain from particle kinematics and the application of Newton's laws of translation and rotation for rigid bodies to explain the three trajectory types when two identical balls collide with equal speed in a groove of variable groove width (figure \ref{fig:chain}). Overall, an increasing groove width causes an increasing center-of-mass and a decreasing tangential acceleration in magnitude, so that when the rolling condition is fulfilled, the center-of-mass velocity either points away from the location of collision (the balls depart), is zero (the balls come to rest) or points towards the location of collision (the balls collide again).

Only when identical balls collide central and ideal-elastic with equal initial speeds, the description of the experiment can be restricted to one ball and the three trajectory types arise independently of the initial speeds. Otherwise, the motions of the balls are asymmetric, depend on the initial speeds and also on the inelasticity of the collision \cite{Domenech}.

The theoretical level of the experiment provides a challenging problem for introductory mechanics courses. Explaining the occurence of the three trajectory types in the experiment requires to make appropriate assumptions, to apply and combine several rigid body concepts and to perform a quantitative analysis of the relations between quantities. Therefore, students should have sufficient previous knowledge about laws and kinematics of rigid bodies from the lecture. We recommend to implement the experiment with the here presented theoretical treatment in homework problems or in the introductory physics laboratory \cite{Hanisch}, because there is sufficient time for students to understand the experimental results on their own.
Additionally, teachers can deal with the well known difficulties students have in understanding relevant rigid body concepts concerning the experiment, for example:
\begin{itemize}[wide]
    \item Particle velocities of a rolling body with respect to a fixed and a moved coordinate system \cite{Rimoldini}.
\item Independence of Newton's law of translation from the working point of forces \cite{Close}.
\item A single force on a rigid body can cause translation and rotation \cite{Close}.
\item Connection of kinetic friction forces and kinematics during the transition from rolling to gliding \cite{Ambrosis}.
\end{itemize}

\bibliography{article} 
\bibliographystyle{iopart-num}

\end{document}